\documentclass[secnumarabic,aps,prd,floatfix,nofootinbib,showkeys,twocolumn,showpacs,10pt,superscriptaddress] 
{revtex4-1}

\usepackage{times}
\usepackage{amsfonts}
\usepackage{amsmath}
\usepackage{amssymb}
\usepackage{natbib}
\usepackage{graphicx}
\usepackage{enumitem}
\usepackage{xcolor}
\usepackage[colorlinks=true,linkcolor=red, citecolor=blue]{hyperref}
\usepackage[outdir=./]{epstopdf}
\usepackage[normalem]{ulem}
\usepackage{amsmath}
\usepackage{booktabs}

\def\aap{A\&A}

\def\apj{ApJ}
\def\apjl{ApJL}

\def\jcap{JCAP}
\def\mnras{MNRAS}
\def\na{New Astronomy}

\begin{document}
\definecolor{orange}{rgb}{0.9,0.45,0}
\def\CovDev{D}
\def\Res{{\mathcal R}}
\def\Gammaflat{\hat \Gamma}
\def\metricflat{\hat \gamma}
\def\Dflat{\hat {\mathcal D}}
\def\part_n{\partial_\perp}
%
\def\Lie{\mathcal{L}}
\def\A{\mathcal{X}}
\def\Aphi{\A_{\phi}}
\def\hAphi{\hat{\A}_{\phi}}
\def\E{\mathcal{E}}
\def\Ham{\mathcal{H}}
\def\M{\mathcal{M}}
\def\R{\mathcal{R}}
\def\p{\partial}
\def\hg{\hat{\gamma}}
\def\hA{\hat{A}}
\def\hD{\hat{D}}
\def\hE{\hat{E}}
\def\hR{\hat{R}}
\def\hcA{\hat{\mathcal{A}}}
\def\hDelt{\hat{\triangle}}
\def\na{\nabla}
\def\dif{{\rm{d}}}
\def\non{\nonumber}
\newcommand{\erf}{\textrm{erf}}
\newcommand{\saeed}[1]{\textcolor{blue}{SF: #1}} 
%
\renewcommand{\t}{\times}
\long\def\symbolfootnote[#1]#2{\begingroup%
\def\thefootnote{\fnsymbol{footnote}}\footnote[#1]{#2}\endgroup}
\title{Toward Gravitational Lensing in Modified Theories of Gravity} 

\author{Ali Tizfahm} 
\email{a.tizfahm@email.kntu.ac.ir}
\affiliation{Department of Physics, K.N. Toosi University of Technology, P.O. Box 15875-4416, Tehran, Iran}
\affiliation{PDAT Laboratory, Department of Physics, K.N. Toosi University of Technology, P.O. Box 15875-4416, Tehran, Iran}

\author{Saeed Fakhry}
\email{s\_fakhry@sbu.ac.ir}
\affiliation{PDAT Laboratory, Department of Physics, K.N. Toosi University of Technology, P.O. Box 15875-4416, Tehran, Iran}
\affiliation{Department of Physics, Shahid Beheshti University, 1983969411, Tehran, Iran}

\author{Javad T. Firouzjaee} 
\email{firouzjaee@kntu.ac.ir}
\affiliation{Department of Physics, K.N. Toosi University of Technology, P.O. Box 15875-4416, Tehran, Iran}
\affiliation{PDAT Laboratory, Department of Physics, K.N. Toosi University of Technology, P.O. Box 15875-4416, Tehran, Iran}

\author{Antonino Del Popolo}
\email{antonino.delpopolo@unict.it}
\affiliation{Dipartimento di Fisica e Astronomia, University of Catania, Viale Andrea Doria 6, 95125 Catania, Italy}
\affiliation{INFN Sezione di Catania, Via Santa Sofia,64, 95123 Catania, Italy}

\date{\today}

\begin{abstract} 
\noindent
In this study, we investigate gravitational lensing within modified gravity frameworks, focusing on the Hu-Sawicki $f(R)$ and normal branch Dvali-Gabadadze-Porrati (nDGP) models, and we compare these results with those obtained from general relativity (GR). Our results reveal that both modified gravity models consistently enhance key lensing parameters relative to GR, including the Einstein radius, lensing optical depth, and time delays. Notably, we find that the Hu-Sawicki $f(R)$ and nDGP models yield significantly larger Einstein radii and higher lensing probabilities, especially at greater redshifts, indicating an increased likelihood of lensing events under modified gravity. Our analysis of time delays further shows that the broader mass distributions in these frameworks lead to pronounced differences in high-mass lens systems, providing potential observational markers of modified gravity. Additionally, we observe amplified magnification factors in wave optics regimes, highlighting the potential for gravitational wave (GW) lensing to differentiate modified gravity effects from GR predictions. Through these findings, we propose modified gravity theories as compelling alternatives to GR in explaining cosmic phenomena, with promising implications for future high-precision gravitational lensing surveys.
\end{abstract}


\maketitle
\vspace{0.8cm}

\section{Introduction} 
Gravitational lensing describes how the path of null waves originating from a distant source is altered as they pass near a massive object. This change in direction results from the gravitational field of the massive lens, which bends the path of these waves in line with general relativity (GR). This phenomenon, affecting both electromagnetic (EM) waves and gravitational waves (GWs), can produce interference patterns and distinct image structures shaped by the characteristics of the lens and the configuration of various potential light paths. Gravitational lensing thus offers a valuable tool in astrophysics and cosmology, influencing how we detect and interpret all forms of radiation from distant sources \citep{1998LRR.....1...12W, 2010CQGra..27w3001B, 2018GReGr..50...42C, 2023FrP....1113909C, 2003ApJ...595.1039T, 2020PhRvD.101f4011H, 2021PhRvD.104j3529D}. 

Lensing of EM waves has already transformed cosmological studies by enabling precise measurements of dark matter distributions, identifying faint celestial objects, and even detecting exoplanets around distant stars \cite{bond2004ogle, massey2010dark, welch2022highly}. Since the landmark 2015 detection of GW event GW150914 by LIGO-Virgo, we now recognize that GWs also undergo lensing, a process which holds the potential to further enhance our cosmological understanding by providing new insights into the distribution of mass in the Universe and the properties of dark matter \cite{maggiore2007gravitational, liao2022strongly}. For example, lensed GWs can address issues such as the mass-sheet degeneracy that limits EM lensing \cite{2021breaking} and open new avenues for testing gravitational theories, such as through the study of GW birefringence and propagation speed \cite{ezquiaga2020gravitational, baker2017multimessenger}.

One of the most profound implications of gravitational lensing lies in its potential to reveal the nature of dark matter \cite{2010RPPh...73h6901M, 2023arXiv230611781V}. In cosmology, dark matter remains an elusive but critical component of the Universe, inferred from gravitational interactions observed in galactic and sub-galactic structures. Studies suggest that sub-galactic dark matter halos could serve as strong lenses for compact background sources, producing characteristic milli-arcsecond scale lensing effects \cite{2022A&A...668A.166L}. By observing these ``milli-lenses," one can test dark matter models, such as cold dark matter halos, whose lensing effects could vary based on the properties of null waves \cite{jung2019gravitational}. Thus, gravitational lensing provides a unique lens through which to investigate dark matter structures and distributions that are otherwise invisible to direct observation. In gravitational lensing research, selecting suitable dark matter halo models is crucial \cite{TIZFAHM2024101712}. Dark matter halo formation scenarios have been discussed in various models, e.g., \cite{2022MNRAS.517L..46W, 2023MNRAS.520.4370D, 2023PDU....4101259D, 2023arXiv231115307F, 2024ARep...68...19D}. In this regard, recent studies have demonstrated that realistic dark matter halo models yield predictions more closely aligned with contemporary cosmological data, see, e.g., \cite{2021PhRvD.103l3014F, 2022PhRvD.105d3525F, 2022ApJ...941...36F, 2023PDU....4101244F, 2023PhRvD.107f3507F, 2023ApJ...947...46F, 2023arXiv230811049F, 2024ApJ...966..235F, 2024arXiv240811995F}.

Yet, while GR has successfully described gravitational lensing, it encounters challenges in accounting for the dark sector of the Universe, where large amounts of unseen dark matter are required to explain observed gravitational effects, see, e.g., \cite{1998CQGra..15..933T, 2023EPJC...83..100S}. As the exact nature of dark matter remains unknown, modified gravity theories have gained attention by offering explanations for these observations without invoking dark matter, see, e.g., \cite{PhysRev.124.925, 2000PhLB..485..208D, 2006JCAP...03..004M, 2010RvMP...82..451S, 2011PhRvD..84b4020H, 2019EPJC...79..708X}. Such theories propose that gravitational behavior might differ on galactic or cosmic scales, suggesting that the apparent need for dark matter could instead reflect limitations in our understanding of gravity. These alternative frameworks provide an exciting path to explore whether modifications to gravity itself could explain lensing phenomena and reveal new aspects of cosmic structure \cite{2008PhRvD..78d3002S, 2011PhRvD..83b4030N, 2019PhLB..795..144G, 2021JCAP...10..062R, 2022PhRvD.106f4012K, 2023JCAP...11..059K, 2024MNRAS.533.3546D}.

In this work, we focus on two widely studied modified gravity models: the Hu-Sawicki $f(R)$ gravity model \cite{2007PhRvD..76f4004H} and the normal branch of the Dvali-Gabadadze-Porrati (nDGP) model \cite{2000PhLB..485..208D}. Both models propose modifications to GR that allow for differences in gravitational effects at large scales. For instance, the Hu-Sawicki model introduces a functional form of the Ricci scalar that mimics GR on local scales but can drive cosmic acceleration without dark energy on larger scales \cite{2007PhRvD..75l7502S, 2012A&A...548A..31S, 2016MNRAS.459.3880H, 2022PhRvD.106j3533M}. In contrast, the nDGP model suggests a braneworld scenario where our Universe exists within a four-dimensional ``brane" embedded in a higher-dimensional space, resulting in gravity that behaves normally on small scales but changes at cosmic distances \cite{2021PhRvD.104j3519L, 2024JCAP...07..093S}. These modified theories predict subtle but measurable deviations from GR in gravitational lensing signals, offering potential tests of their validity. For example, \( f(R) \) gravity could lead to enhanced lensing effects, while the nDGP model could yield scale-dependent changes in lensing across different distances.

In this work, we examine gravitational lensing under the Hu-Sawicki \( f(R) \) and nDGP models, analyzing how each theory impacts key observables such as Einstien radius, lensing strength, and time delays in strong lensing systems. By comparing these predictions to GR, we aim to identify distinct signatures that could be detected by next-generation observational tools, such as advanced GW and EM lensing observatories \cite{2024arXiv240513491E, 2024arXiv240608919S}. These efforts could ultimately clarify whether GR or modified gravity better explains the cosmos, potentially advancing our understanding of the Universe’s structure.

The work is structured as follows: Section II presents the theoretical foundations of the Hu-Sawicki \( f(R) \) and nDGP models, highlighting key distinctions from GR. Section III describes our methods for calculating lensing properties in these modified gravity frameworks, including screening mechanisms to limit effects in high-density regions. Section IV details our findings, and Section V concludes with a summary and suggestions for future research.
 
\section{Modified Gravity Models}
In this section, we discuss two frequently studied modified gravity models: the Hu-Sawicki $f(R)$ model and the nDGP model. These models illustrate the screening mechanisms known as the Chameleon and Vainshtein mechanisms, respectively.

\subsection{Hu-Sawicki $f(R)$ Model}
The Hu-Sawicki $f(R)$ model introduces a non-linear modification function, denoted as $f(R)$, into the conventional Einstein-Hilbert action \citep{2007PhRvD..76f4004H}:
\begin{equation}\label{frgravity}
S=\int d^{4}x \sqrt{-g}\left[\frac{R+f(R)}{2\kappa}+\mathcal{L}_{\rm m}\right],
\end{equation}
where $R$ represents the Ricci scalar, $\kappa$ is the Einstein gravitational constant, $g$ is the determinant of the metric, and $\mathcal{L}_{\rm m}$ refers to the matter Lagrangian. By setting $f = -2\Lambda$, where $\Lambda$ is the cosmological constant, GR can be recovered.

Using a conformal transformation, Eq.\,(\ref{frgravity}) can be rewritten as a scalar-tensor theory that includes a scalar field, denoted as $f_{\rm R} \equiv {\rm d}f(R)/{\rm d}R$, representing the extra degree of freedom introduced by the modification. The functional form of $f(R)$ must be chosen to drive late-time cosmic acceleration while complying with solar system tests \citep{2008PhRvD..78j4021B}. A widely studied class of models achieving this is the Hu-Sawicki model \citep{2007PhRvD..76f4004H}, described by:
\begin{equation}
f_{R}=-m^2\frac{c_1(R/m^2)^n}{c_2(R/m^2)^n+1},
\end{equation}
where $m^2=\kappa\bar{\rho}_{\rm{m}0}/3$ is the characteristic mass scale, and $\bar{\rho}_{\rm{m}0}$ is the present-day background matter density. The dimensionless parameters $c_1$, $c_2$, and $n>0$ are free parameters that must be carefully selected to match the expansion history and satisfy solar system constraints through the chameleon mechanism.

Ensuring stability in high-density regions, where $R \gg m^2$, is critical. Additionally, cosmological tests of the $f(R)$ model should align with GR-based observations. To satisfy this condition, the second derivative of $f(R)$ with respect to $R$, denoted $f_{RR} = {\rm d}^2 f/{\rm d}R^{2}$, must be positive. Consequently, the Hu-Sawicki model can be approximated as follows:
\begin{equation}\label{limfr}
\lim_{m^2/R \rightarrow 0} f(R) \approx -\frac{c_1}{c_2}m^2+\frac{c_1}{c_2^2}m^2\left(\frac{m^2}{R}\right)^n.
\end{equation}

Although this model does not explicitly include a cosmological constant, it behaves similarly in large-scale and local observations. Moreover, the finite ratio $c_1/c_2$ results in a constant curvature that is independent of changes in the matter density. Thus, this class of models accelerates the Universe's expansion, mimicking the behavior of the standard cosmological model. The relation in Eq.\,(\ref{limfr}) can be further simplified as:
\begin{equation}
f(R) \approx -\frac{c_1}{c_2}m^2-\frac{f_{R_{0}}}{n}\frac{\bar{R}_{0}^{n+1}}{R^n},
\end{equation}
where $\bar{R}_{0}$ indicates the present-day background curvature, and $f_{R0} \equiv f_R(\bar{R}_0)$ is the field strength. If $|f_{R0}| \rightarrow 0$, $(c_1/c_2) m^2 = 2\kappa \bar{\rho}_{\Lambda}$ can be obtained, where $\bar{\rho}_{\Lambda}$ represents the background energy density attributed to dark energy. Cosmological and solar system tests have restricted the field strength $f_{R0}$ \citep{2020PhRvD.102j4060D}. Various studies have explored $|f_{R0}|$ values between $10^{-4}$ and $10^{-8}$ \citep{2019JCAP...09..066M}. In this work, we investigate the Hu-Sawicki model for $n = 1$ and $|f_{R0}| = 10^{-4}$, $10^{-5}$, and $10^{-6}$, referred to as $f4$, $f5$, and $f6$, respectively.

\subsection{nDGP Model}
The nDGP model of gravity is a MG framework proposed by \cite{2000PhLB..485..208D}. It describes the Universe as a four-dimensional brane embedded in a five-dimensional Minkowski space. The action consists of two terms:
\begin{equation}\label{ndgpgravity}
S=\int d^{4}x \sqrt{-g}\left[\frac{R}{2\kappa}+\mathcal{L}_{\rm m}\right] + \int d^{5}x \sqrt{-g_5}\frac{R_{5}}{2\kappa_{5} r_{\rm c}},
\end{equation}
where $R_5$, $g_5$, and $\kappa_5$ are the Ricci scalar, metric determinant, and Einstein gravitational constant for the fifth dimension, respectively. The crossover distance $r_{\rm c}=(\kappa_5/2 \kappa)$ represents the scale at which GR is valid in four dimensions. For distances larger than $r_{\rm c}$, the second term in the action becomes significant, causing deviations from GR.

The DGP model includes two branches: the "normal" branch (nDGP) and the "self-accelerating" branch (sDGP). We focus on the nDGP branch, as it avoids ghost instabilities \citep{2009PhRvD..80f3536L}. On large scales, gravity strengthens, while at small scales, it resembles GR due to Vainshtein screening. This makes the nDGP model capable of replicating the $\Lambda$CDM expansion history, a key advantage supported by numerous simulations. In this model, the sole parameter to be constrained is $n = H_0 r_{\rm c}$ (where $H_0$ is the Hubble constant), with values between $1$ and $5$ commonly explored. GR is recovered as $n \rightarrow \infty$, corresponding to a steep gradient in gravitational strength under Vainshtein screening. Several studies have examined the implications of the nDGP model for structure formation and cosmology through numerical simulations and observational data. In this work, we explore the nDGP model with $n=1$, $2$, and $5$, referred to as nDGP(1), nDGP(2), and nDGP(5), respectively.

\section{Dark Matter Halo Models}\label{sec:iii}
This section offers a concise review of dark matter halo models derived from modified gravity (MG) frameworks. These models are pivotal as initial conditions in estimating the key parameters of gravitational lensing in MG models.

\subsection{Halo Density Profile}
In the context of cosmological perturbation theory, dark matter halos represent dynamic systems in the nonlinear regime. These halos are described by their density distribution, which is a function of radius, commonly referred to as the halo density profile. Essentially, a halo density profile characterizes how dark matter is distributed within a galactic halo, and serves as a predictive tool for the dark matter distribution in host halos.

Over time, various analytical methods and numerical simulations have been employed to derive density profiles that accurately correspond to observed data, particularly the rotation curves of galaxies, see, e.g., \cite{1965TrAlm...5...87E, 1996ApJ...462..563N}. The Navarro-Frenk-White (NFW) profile is among the most widely applied dark matter halo density profiles \citep{1996ApJ...462..563N}
\begin{equation}\label{nfw}
\rho(r)=\frac{\rho_{\rm s}}{r/r_{\rm s}(1+r/r_{\rm s})^2},
\end{equation}
where $\rho_{\rm s}$ and $r_{\rm s}$ are scale density and radius. The NFW profile employs a two-parameter function to describe the halo mass distribution as a function of distance from the halo center.  For the NFW profile, the logarithmic slope of the density distribution is $-2$ at the scale radius, $r_{\rm s}$.

\subsection{Halo Concentration Parameter}
The halo concentration parameter is a dimensionless quantity representing the core density of galactic halos, defined as the ratio of the halo's virial radius, $r_{\rm vir}$, to  $r_{\rm s}$. The virial radius defines a region where the average density is $200$ to $500$ times the critical density of the Universe. This parameter is crucial in cosmology as it informs the shape of the halo density profile and the distribution of subhalos. Simulations and theoretical studies suggest that the concentration parameter evolves dynamically with halo mass and redshift \cite{2012MNRAS.423.3018P, 2014MNRAS.441.3359D, 2016MNRAS.460.1214L, 2016MNRAS.456.3068O}. This reflects the correlation between halo mass and merger history, with smaller halos having completed virialization and thus being more concentrated than their host halos. For this study, we employ the concentration parameter defined in \cite{2016MNRAS.456.3068O} within a general relativistic (GR) framework, and also use the parameters derived in \cite{2019MNRAS.487.1410M} and \cite{2021MNRAS.508.4140M} for the Hu-Sawicki $f(R)$ gravity model and the nDGP model, respectively.

\subsection{Halo Mass Function}
The halo mass function is a crucial concept for categorizing halos based on their mass, providing valuable insight into the distribution of dark matter halos as a function of mass. The density contrast is expressed as $\delta(x) \equiv [\rho(x)-\bar{\rho}]/\bar{\rho}$, where $\rho(x)$ is the density at point $x$, and $\bar{\rho}$ is the background mean density. The halo mass function serves as a key tool for identifying cosmological structures by mass, particularly those that exceed a certain density contrast threshold, leading to their collapse and subsequent halo formation.

Predicting the halo mass function is a vital test for cosmological models. The Press-Schechter (PS) approach offers an analytical framework to model the hierarchical structure formation statistics from initial density perturbations \citep{1974ApJ...187..425P}. In this framework, dark matter halo formation is viewed as a stochastic process involving Gaussian density fields, which are smoothed by a filtering function. Although the PS formalism is widely used, it often underestimates halo abundances. Thus, various refinements have been developed to address this limitation \footnote{The Sheth-Tormen (ST) formalism is a notable improvement over the PS model, as it incorporates triaxial collapse \citep{2001MNRAS.323....1S}. The ST model provides better alignment with simulations but still overestimates high-mass halos. Despite refinements that improve its accuracy, discrepancies remain, particularly at high redshifts when compared to Bolshoi simulations \citep{2011ApJ...740..102K}.}.

The PS model was further refined in \cite{1998A&A...337...96D}, which showed that the collapse threshold depends on mass, rather than being constant. As a result, the collapse threshold can be expressed as:
\begin{equation} \label{eqn:barrier}
\delta_{\rm cm} = \delta _{\rm c}(z) \left(1+\frac{\beta}{\nu^\alpha}\right),
\end{equation}
where $\alpha=0.585$ and $\beta=0.46$. Here, $\nu=(\delta_{\rm c}/\sigma)^{2}$, with $\sigma(M, z)$ representing the linear root-mean-square fluctuation of overdensities on a comoving scale of mass $M$ at redshift $z$. Using excursion set theory, it can be shown that Eq.\,(\ref{eqn:barrier}) leads to the barrier:
\begin{equation}\label{eqn:barrierv}
B(M)=\sqrt{a} \delta _{\rm c}(z) \left(1+\frac{\beta}{a \nu^\alpha} \right).
\end{equation}
With an appropriate choice of $a$, this barrier produces a mass function consistent with simulations \citep{1998A&A...337...96D}. Modifications to the barrier can include additional physical effects like mergers, fragmentation, tidal torques, and the cosmological constant, as shown in \cite{2017JCAP...03..032D}, leading to a more refined model for halo formation:
\begin{eqnarray}\label{eqn:barrierf}
\delta_{\rm cm2} = \delta_{\rm co} \left[1+\frac{\beta}{\nu^{\alpha}}+\frac{\Omega_{\Lambda}\beta_2}{\nu^{\alpha_2}}+\frac{\beta_3} {\nu^{\alpha_3}}\right],
\end{eqnarray}
where $\alpha=0.585$, $\beta=0.46$, $\alpha_2=0.4$, $\beta_2=0.02$, $\alpha_3=0.45$, and $\beta_3=0.29$. Here, $\Omega_{\Lambda} \simeq 0.692$ is the cosmological constant density parameter.

In the excursion set theory framework, the mass function is defined as the comoving number density of halos within a given mass range \citep{1991ApJ...379..440B}:
\begin{equation}\label{eqn:universal}
n(M,z)=\frac{\overline{\rho}}{M^{2}}\left|\frac{d\log{\nu }}{d\log M}\right|\nu f(\nu)\;,
\end{equation}
where $f(\nu)$ is the multiplicity function, describing the distribution of first-crossing.

The following relation can be obtained by analyzing the barrier given in Eq.\,(\ref{eqn:barrierf}) and factoring in the effects of angular momentum, dynamical friction, and the cosmological constant, as illustrated in \cite{2017JCAP...03..032D}:
\begin{equation}\label{nufnu}
\nu f(\nu) \simeq A_{2}\sqrt{\frac{a\nu}{2\pi}}l(\nu)\exp{\left\{-0.4019 a \nu^{2.12} m(\nu)^{2}\right\}},
\end{equation}
where
\begin{eqnarray}
& l(\nu)=\left(1+\dfrac{0.1218}{\left(a\nu\right)^{0.585}}+\dfrac{0.0079}{\left(a\nu\right)^{0.4}}+\dfrac{0.1}{\left(a\nu\right)^{0.45}}\right),\nonumber\\~\\
& m(\nu)=\left(1+\dfrac{0.5526}{\left( a\nu \right)^{0.585}} + \dfrac{0.02}{\left(a\nu\right)^{0.4}}+\dfrac{0.07}{\left(a\nu\right)^{0.45}}\right)\nonumber.
\end{eqnarray}
The constants are defined with specific values, namely $A_2=0.93702$ and $a=0.707$. By substituting Eq.\,(\ref{nufnu}) into Eq.\,(\ref{eqn:universal}), one can derive an accurate semianalytical mass function, referred to hereafter as the mass function for GR.

However, in modified gravity theories, rescaling the mass function is essential. This is because, in such models that introduce an extra scalar degree of freedom, Birkhoff’s theorem does not hold. As a result, the collapse process becomes dependent on environmental factors, a phenomenon known as the chameleon screening mechanism. In this mechanism, dense regions partially shield against the enhanced gravitational forces. Consequently, the threshold for over-dense regions is influenced by the halo mass, redshift, and specific model parameters, which affect the gradients of the scalar field.

In \cite{2022PhRvD.105d3538G}, a rescaled halo mass function for Hu-Sawicki $f(R)$ gravity is presented based on $N$-body simulations:
\begin{equation}\label{massfuncfr}
\frac{n_{f(R)}}{n_{\rm GR}}= 1 + A\exp\left[\frac{\left(X-B\right)^{2}}{C^{2}}\right],
\end{equation}
where $X\equiv\ln(\sigma^{-1})$, which naturally depends on $M_{\rm vir}$ and $z$. Additionally, based on various Hu-Sawicki $f(R)$ models, the optimal values of $A$, $B$, and $C$ are provided in Table \ref{tab1}.

In \cite{2021MNRAS.508.4140M}, a suitable form for the halo mass function in nDGP models is proposed, derived from dark matter-only simulations:
\begin{eqnarray}\label{massfuncndgp}
\frac{n_{\rm nDGP}}{n_{\rm GR}}= 1 + D(H_{0}r_{\rm c})\times\hspace*{3.5cm}\nonumber\\
\left[\tanh\left(\log(M_{\rm vir}/M_{\odot}h^{-1})-E(z)\right)+F(z)\right],
\end{eqnarray}
where
\begin{eqnarray}
& D(H_{0}r_{\rm c}) = (0.342 \pm 0.014)(H_{0}r_{\rm c})^{-1}, 
\end{eqnarray}
\begin{eqnarray}
& E(z) = (14.87 \pm 0.03) - (0.481 \pm 0.01)z, 
\end{eqnarray}
\begin{eqnarray}\label{end}
& F(z) = (0.864 \pm 0.008) + (0.047 \pm 0.005)z. 
\end{eqnarray}
It is worth noting that these parameters represent the best-fit values obtained through the method of unweighted least squares.

\begin{table}
\caption{Best-fit values of parameters in Eq.\,(\ref{massfuncfr}) for the Hu-Sawicki $f(R)$ models \citep{2022PhRvD.105d3538G}.}
\begin{tabular}{|c|c|c|c|}
\hline
\hline
Model & A & B & C \\
\hline
\hline
f4 & 0.630 & 1.062 & 0.762 \\
\hline
f5 & 0.230 & 0.100 & 0.360 \\
\hline
f6 & 0.152 & -0.583 & 0.375 \\
\hline
\hline
\end{tabular}
\label{tab1}
\end{table}

\section{Strong Gravitational Lensing}\label{sec:iii}
In this section we will investigate the basic concepts of gravitational lensing including, Einstein radius, halo velocity dispersion, lensing probability and time delay. 
\subsection{Basic Lens Equations}
Gravitational lensing of GWs offers distinct advantages over EM spectrum methods, providing a direct approach to probing the gravitational potential and mass distribution of astrophysical lenses \citep{2023PhRvD.108d3527T}. While most studies on cold dark matter (CDM) subhalos through gravitational lensing focus on subhalos beyond the Local Group (typically at redshifts around $\sim 0.5 - 1.0$), subhalos within the Milky Way’s dark matter halo can also induce lensing effects \cite{2010AdAst2010E...9Z}. However, detecting and distinguishing these local effects from other phenomena is quite challenging. In contrast, the oscillatory nature of GWs enables unique wave optics effects, such as diffraction and interference, to manifest during lensing in ways not observed with EM waves. Observable phenomena, such as time delays between multiple lensed images, can provide precise information about lens masses and cosmological parameters, avoiding issues like the mass-sheet degeneracy that complicates traditional light-based lensing studies \citep{2013A&A...559A..37S}. Additionally, GW sources, such as compact binary mergers, act as bright, point-like sources, simplifying the calculation of lensing effects. Their transient nature also allows tracking of time variability in lensing phenomena, making GWs an invaluable tool for studying cosmic structures \citep{2018MNRAS.476.2220L}.

Strong lensing, however, is a relatively rare event, occurring only when light from a distant object passes through or near high-density regions like galaxy centers or clusters. The probability of such strong lensing events can be estimated if the mass distribution and abundance of potential lensing objects are well understood.

The deflection angle of a light ray passing by a spherical body of mass $M$ at a distance $b$ is given by \cite{1992grle.book.....S}.
\begin{equation}
\hat{\alpha}=\frac{4GM}{c^{2} b},
\end{equation}
where $G$ is the gravitational constant and $c$ is the speed of light.

Gravitational lensing can be understood using the principles of geometrical optics. The lens equation, which relates the position of the image to that of the source, can be readily derived from the geometry of the lensing system. This lens equation is:
\begin{equation}
    \beta = \theta - \frac{D_{\rm ls}}{D_{\rm s}} \hat{\alpha},
\end{equation}
This lens equation yields two solutions, corresponding to the locations of the two images produced by the gravitational lensing effect
\begin{equation}
    \theta_{\pm} = \frac{\beta}{2} \pm \frac{\beta}{2} \sqrt{1+\frac{4\theta_{\rm E}^2}{\beta^2}},
\end{equation}
where the positive and negative signs represent the first and second images, respectively. Generally, $\theta_{+}$ denotes the primary image, located on the same side of the lens as the source, while $\theta_{-}$ represents the secondary image, positioned on the opposite side of the lens from the source.

A unique situation occurs when the source is positioned directly behind the lens, meaning $\beta = 0$. Due to the system's rotational symmetry, this alignment results in a ring-shaped image, known as an Einstein ring. The angular radius of this ring, referred to as the Einstein radius, is defined as follows \citep{2006JCAP...01..023M}
\begin{eqnarray}\label{ER1}
\theta_{\rm E}(M)=\sqrt{\dfrac{4 G M}{c^2}\dfrac{D_{\rm ls}}{D_{\rm l}D_{\rm s}}} \hspace*{3.3cm}\nonumber \hspace*{0.5cm}\\
\approx 3 \times 10^{-6}\left(\frac{M}{M_{\odot}}\right)^{1/2} \left[\frac{D_{\rm ls}/(D_{\rm l}D_{\rm s})}{1 {\rm Gpc}}\right]^{1/2} \,\mathrm{arcsec},
\end{eqnarray}
where $D_{\rm s}$, $D_{\rm ls}$, and $D_{\rm l}$ represent the angular-diameter distances from the observer to the source, from the lens to the source, and from the observer to the lens, respectively.

\begin{figure*}\label{Radi}
   \centering
     \includegraphics[width=0.9\linewidth]{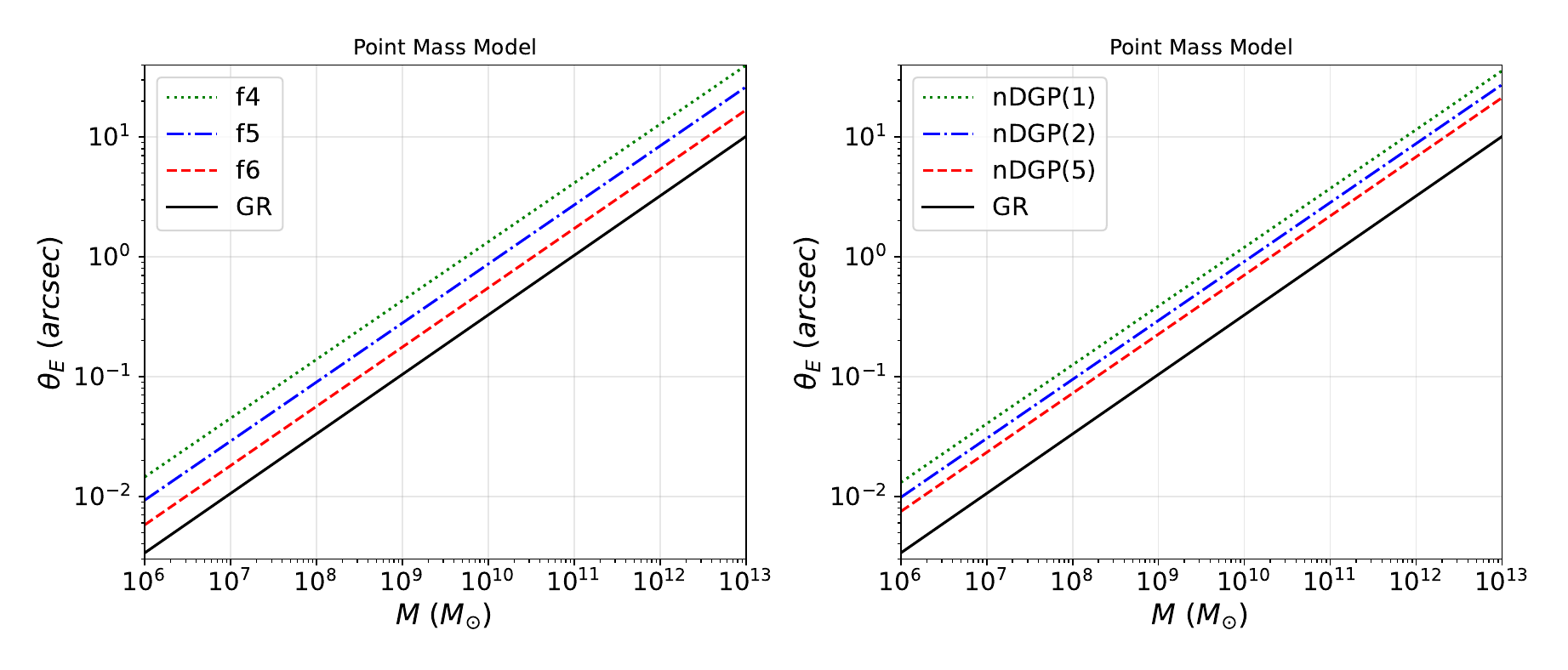}
     \includegraphics[width=0.9\linewidth]{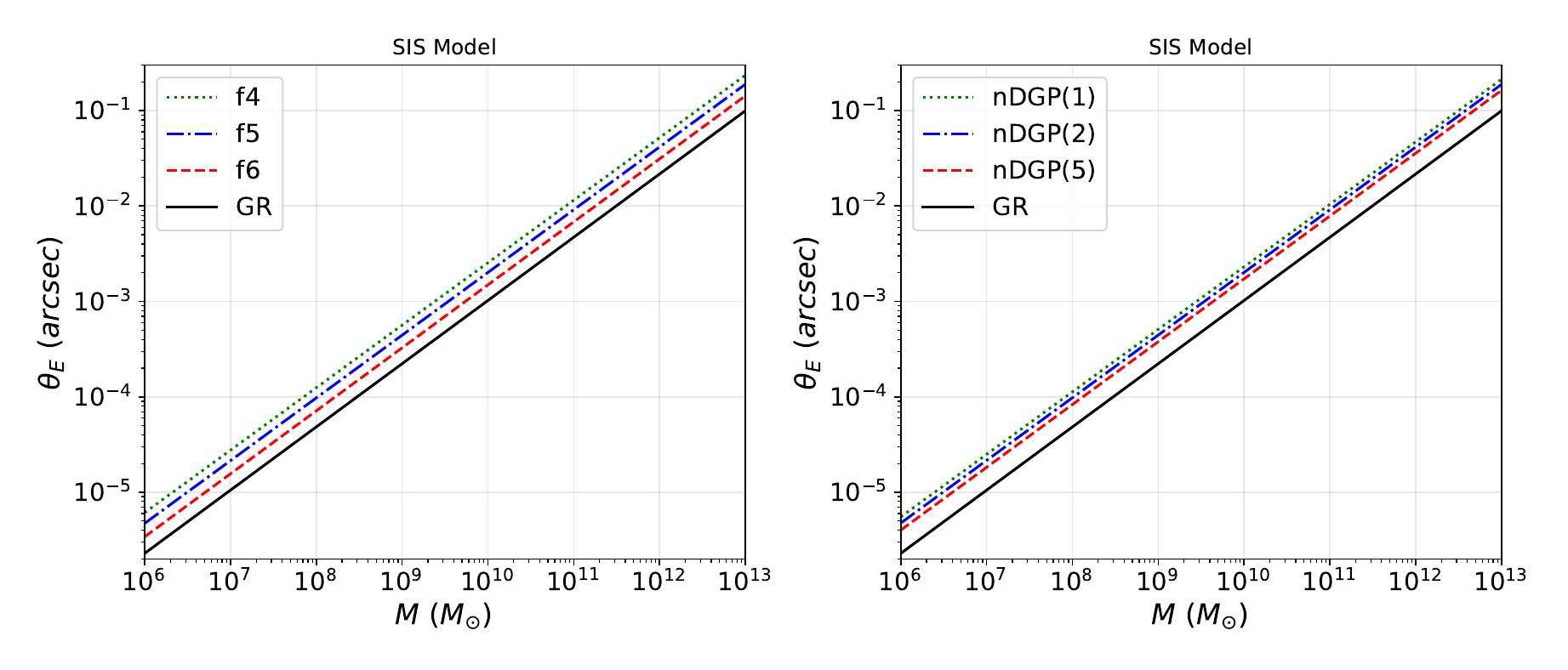}
   \caption{Einstein radii as a function of lens mass for the Hu-Sawicki $f(R)$ (left) and nDGP (right) models, compared to findings in the standard GR framework. The analysis includes both point mass (top) and SIS (bottom) lens models.}
   \label{Fig1}
\end{figure*}

Beyond point-mass lenses, which model compact objects like black holes as lenses, there are more realistic models where galaxies, star clusters, and dark matter halos act as the lensing mediums. These are often described using the singular isothermal sphere (SIS) lens model. In the SIS model, the surface density is characterized by the velocity dispersion $v$ and is expressed as $\Sigma(\xi) = v^{2}/\xi$. In contrast, the surface mass density for a point-mass model is given by $\Sigma(\xi) = M_{\rm l}\delta^2$. The mass within the Einstein radius, denoted as $M_{\rm lz}$, can be defined as $M_{\rm lz}= \frac{4\pi^2 v^4 (1 + z_{\rm l}) D_{\rm ls}}{D_{\rm l} D_{\rm s}}$. Accordingly, the angular Einstein radius in the SIS model is given by:
\begin{eqnarray}\label{ER2}
    \theta_{\rm E}(M)=\sqrt{\frac{16G\pi^2 \sigma^4}{c^2}\frac{D_{\rm ls}^2}{D_{\rm l}^2D_{\rm s}^2}} \hspace*{2cm}\nonumber \hspace*{0.5cm}\\
    \approx 3 \times 10^{-5}\left(\frac{\sigma}{1{\rm km/s}}\right)^{2}\left(\frac{D_{\rm ls}}{D_{\rm s}D_{\rm l}}\right) \,\mathrm{arcsec}.
\end{eqnarray}
In the above equation, $\sigma$ represents the velocity dispersion of particles within the virial radius $r_{\rm vir}$ \cite{2016MNRAS.456.3068O}:
\begin{eqnarray}
    \sigma=\sigma_{15}\left(\frac{M}{10^{15}M_{\odot}h^{-1}}\right)^{1/3} \hspace*{1cm}\mathrm{km/s}, 
\end{eqnarray}
where $\sigma_{15}$ illustrates a constant in the original ellipsoidal-collapse model. For the NFW profile, $\sigma_{15}$ is
\begin{equation}
\sigma_{15} \approx 241 \ln(y^2) + 295 \ln(y) + 1156.
\end{equation}
In the later equation,
\begin{equation}
y=\frac{0.42+0.20 \nu^{-1.23} \pm 0.083 \nu^{-0.6}}{(Ht)^{2/3}},
\end{equation}
where $H$ is the Hubble constant, $t$ represents the collapse time, and $b_{i}$'s are second-order polynomials of the shape parameter $\alpha$. For $0.1 < \alpha < 0.52$, the polynomials can be determined as follows \citep{2016MNRAS.456.3068O}:
\begin{eqnarray}
b_{1}(\alpha) = -173\alpha^{2} + 237\alpha - 14, \hspace*{1cm} \nonumber \\~ \nonumber \\
b_{2}(\alpha) = -389\alpha^{2} -378\alpha +287, \hspace*{1cm} \nonumber \\~ \nonumber \\
b_{3}(\alpha) = 1540\alpha^{2} - 195\alpha +244, \hspace*{1cm} \nonumber \\~ \nonumber \\
b_{4}(\alpha) = 71\alpha^{2} -287\alpha +1205. \hspace*{1cm} 
\end{eqnarray}

\begin{figure*}
   \centering
     \includegraphics[width=0.9\linewidth]{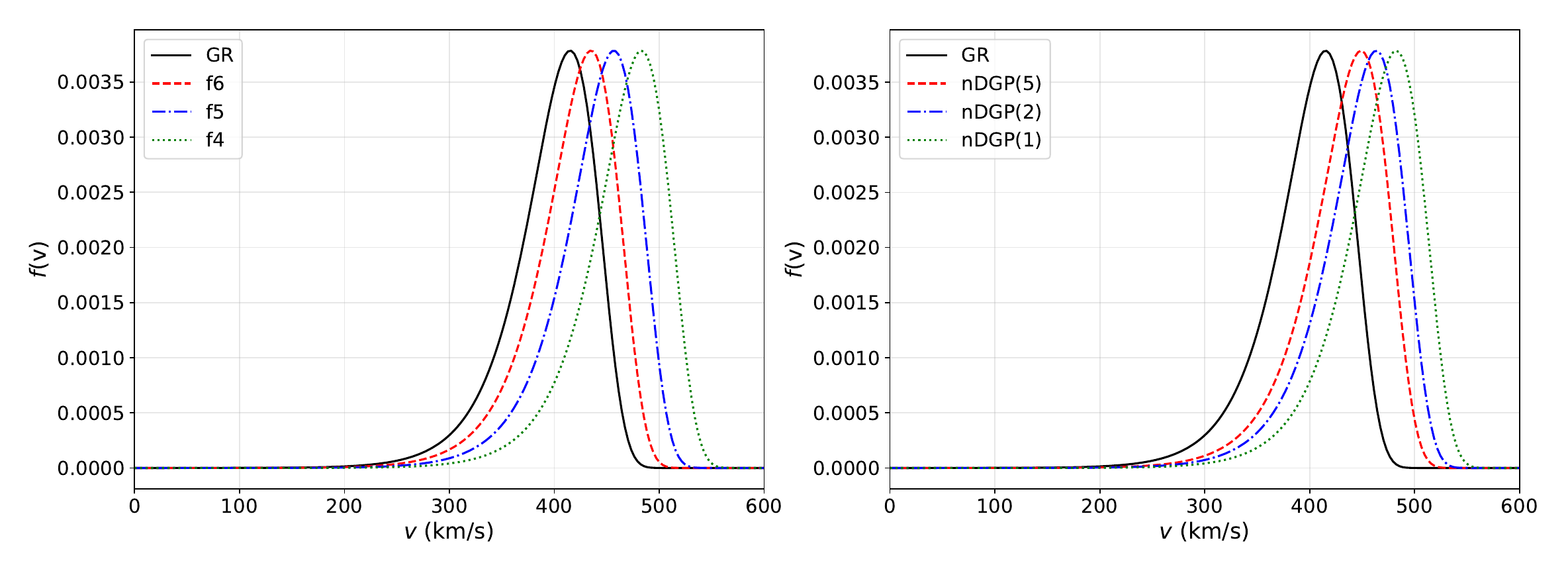}
   \caption{Velocity distribution function of dark matter halos for the Hu-Sawicki $f(R)$ (left) and nDGP (right) models, compared to findings in the standard GR framework while considering the NFW density profile.}
   \label{Fig2}
\end{figure*}

In Fig.\,\ref{Fig1}, we have shown the Einstein radius for point mass and SIS lens models as a function of halo mass, In such calculations, we have considered Hu-Sawicki $f(R)$ and nDGP gravities, and compared them with that obtained from GR. The general trend observed is that the Einstein radius monotonically increases as a function of the associated masses of the dark matter halos. This is consistent with the expectation that more massive lenses will exhibit larger gravitational deflection of light, resulting in larger Einstein radii.

In the upper left panel, the figure compares the results for the Hu-Sawicki $f(R)$ gravity models against the standard GR framework, while considering the point mass lens model. Here, f4, f5, and f6 models exhibits the highest angular Einstein radius compared to that obtained from GR, respectively. This suggests that gravitational lensing effects may be more pronounced in the context of modified gravity models, potentially leading to stronger magnification and image distortion compared to GR. The upper right panel presents a similar comparison, this time for the nDGP models again compared with the GR model. Similar to the f(R) gravity models, nDGP(1), nDGP(2), and nDGP(5) show the highest angular Einstein radius among the models considered compared with GR.

Moreover, in the lower panels, we have depicted the same analysis, but this time for the extended SIS lens model, rather than the point mass lens considered in the upper panels. While the general trends remain similar, with the Einstein radius increasing with halo mass, the SIS model exhibits significantly smaller orders of magnitude for the Einstein radius compared to the point mass lens. This difference can be attributed to the more extended mass distribution of the SIS lens, which has a less concentrated gravitational field and thus a weaker influence on the bending of light, in contrast to the highly localized gravitational potential of the point mass lens. Across both the Hu-Sawicki $f(R)$ gravity and nDGP frameworks, the GR model consistently demonstrates the lowest angular Einstein radius, suggesting that gravitational lensing effects may be more pronounced in the context of modified gravity models compared to that obtained from GR.
\subsection{Velocity Dispersion}
The velocities of dark matter particles within a halo can follow a Maxwellian distribution, meaning they are distributed according to the Maxwell-Boltzmann distribution, which characterizes the statistical behavior of particles in thermal equilibrium. Velocity dispersion can also provide insights into the kinetic energy and temperature of particles within the halo \cite{2012JPhCS.375a2048B}, which is defined as follows:

\begin{equation}\label{veld}
f(\sigma)=\frac{1}{\sqrt{2\pi \lambda}} \exp\left(-\frac{\sigma^2}{2\lambda^2}\right).
\end{equation}
In the above relation, $\lambda \approx 270 \, \text{km/s}$. This velocity distribution corresponds to a simplified model often used as a starting point: an isotropic sphere with uniform temperature and a density profile that decreases with the square of the distance from the center, $\rho (r) \propto r^{-2}$. While this model is advantageous for its simplicity, it does not fully capture the complexity of the halo's density and velocity distribution. Observations and numerical simulations suggest that dark matter halos deviate from this $1/r^2$ density profile, exhibiting more complex structures, such as triaxial shapes and variations in velocity distribution along different axes.

When the velocity distribution is isotropic, a direct relationship exists between $f(v)$ and the spherically symmetric density profile, as described by Eddington's formula \cite{1916MNRAS..76..572E}. Generally, the steady-state phase-space distribution of a collection of collisionless particles is governed by the collisionless Boltzmann equation, while the system's velocity dispersions can be determined using the Jeans equations.

\begin{figure*}
   \centering
     \includegraphics[width=0.9\linewidth]{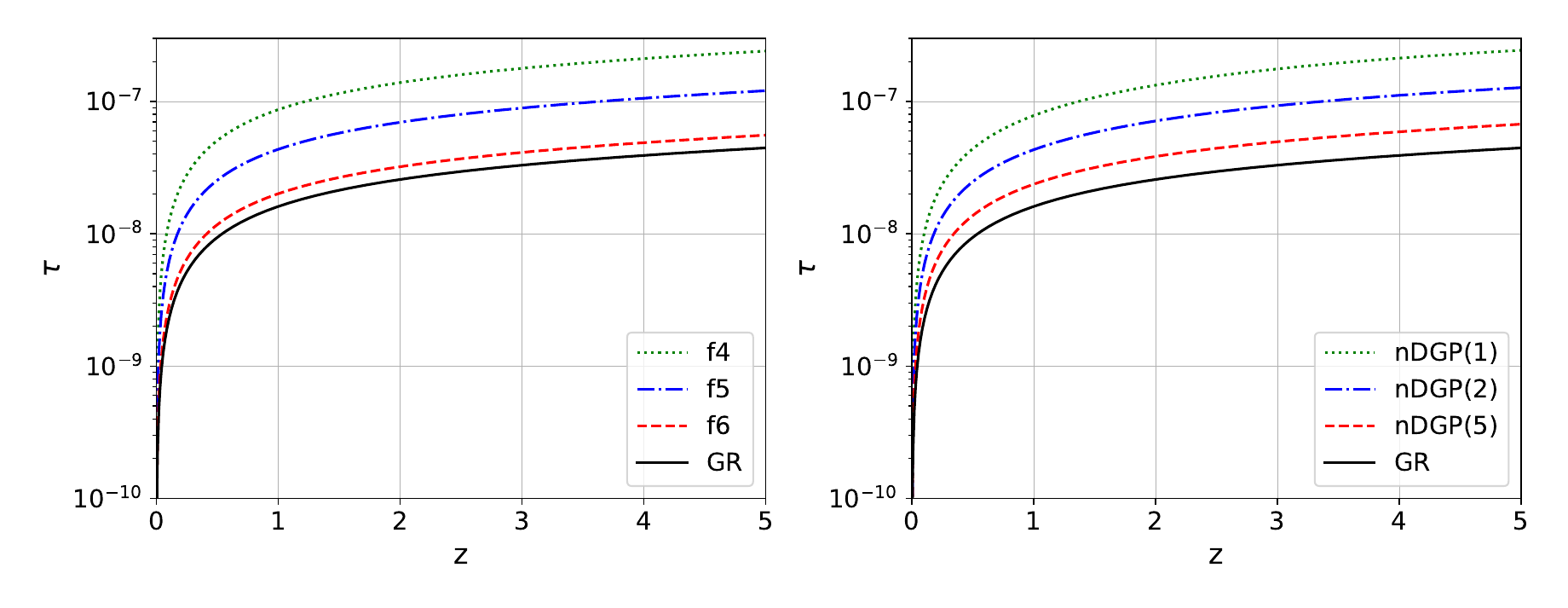}
        \caption{Lensing optical depth as a function of redshift for the Hu-Sawicki $f(R)$ (left) and nDGP (right) models, compared to findings in the standard GR framework while considering the NFW density profile.}
   \label{Fig3}
\end{figure*}

In Fig.\,\ref{Fig2}, we illustrate the velocity distribution function of dark matter particles across various halo models analyzed in this study, including both the GR framework and the modified gravity theories of Hu-Sawicki $f(R)$ gravity and nDGP.

In the left panel, we show the velocity distribution for the GR model alongside three Hu-Sawicki $f(R)$ gravity models (f4, f5, and f6), all using the NFW density profile for dark matter halos. The curves reveal distinct variations among the models, highlighting how the choice of gravitational framework significantly impacts the velocity distribution of dark matter particles. The peaks of these models differ notably, indicating that the underlying gravitational theory influences the kinetic energy and temperature of the particles. Furthermore, at any given value of the velocity distribution function, the range of velocities varies widely between the models, with the f4, f5, and f6 models exhibiting a broader velocity range than the GR scenario.

The right panel presents a comparison between the three nDGP models, nDGP(1), nDGP(2), and nDGP(3), and GR, again using the NFW density profile. Similar to the $f(R)$ gravity models, the velocity distribution curves for the nDGP halos show clear deviations from the GR case, emphasizing the modified gravity framework's impact on the kinetic properties of dark matter. Notably, the nDGP(1), nDGP(2), and nDGP(3) models display a wider velocity range compared to GR.

These results highlight the critical importance of accurately modeling the gravitational framework when examining the velocity distributions and associated kinetic properties of dark matter particles in galactic halos. The observed variations between the GR and modified gravity models suggest that the choice of gravitational theory can profoundly influence our understanding of dark matter dynamics on both galactic and cosmological scales.
\subsection{Lensing Probability}
The likelihood that a source at redshift \( z_{s} \) will be gravitationally lensed by an intervening mass distribution is described by a Poisson-like distribution, expressed as  
\begin{equation}  
 P = 1 - e^{-\tau(z_{\rm s})},  
\end{equation}  
where \( \tau(z_{\rm s}) \) is called the lensing optical depth, which denotes the integrated surface mass density of potential lensing masses along the line of sight to the source.

To estimate the number of gravitationally lensed sources observed at a specific redshift, it is essential to compute the lensing optical depth, which depends on the distance to the sources. The lensing optical depth measures the total probability of a source being lensed by the matter distribution along its line of sight, accounting for the cross-sectional area and number density of potential lenses, such as galaxies and clusters, encountered by the light from the distant source to the observer \citep{2000ApJ...531..613B, 2006A&A...447..419F, 2023JCAP...07..007F}. We assume that lensing cross sections can be computed by treating the lens as an external field. The cross section for lensing by dark matter halos is calculated under the assumption that their density distributions can be modeled as SIS profiles. The lensing optical depth has the form  
\begin{equation}  
\tau(z_{\rm s})=\int_0^{z_{\rm s}} \left|\frac{cdt}{dz}\right|dz \int \zeta\left(M_{\rm l},z\right) \frac{dn\left(M_{\rm l},z\right)}{dM_{\rm l}} dM_{\rm l},  
\end{equation}  
where \( \zeta(M_{\rm l},z) \) denotes the lensing cross-section, defined as follows:  
\begin{equation}  
\zeta(M_{\rm l},z)=\pi \theta_{\rm E}^2 D_{\rm l}^2,  
\end{equation}  
and  
\begin{equation}  
\frac{dt}{dz}=-\frac{1}{H(1+z)}.  
\end{equation}  
Additionally, \( {dn(M_{\rm l},z)}/{dM_{\rm l}} \) describes how the number density of dark matter halos is distributed across different mass ranges, according to various dark matter halo models (see Eqs.\,(\ref{eqn:universal} - \ref{end})).

\begin{figure*}
   \centering
     \includegraphics[width=0.9\linewidth]{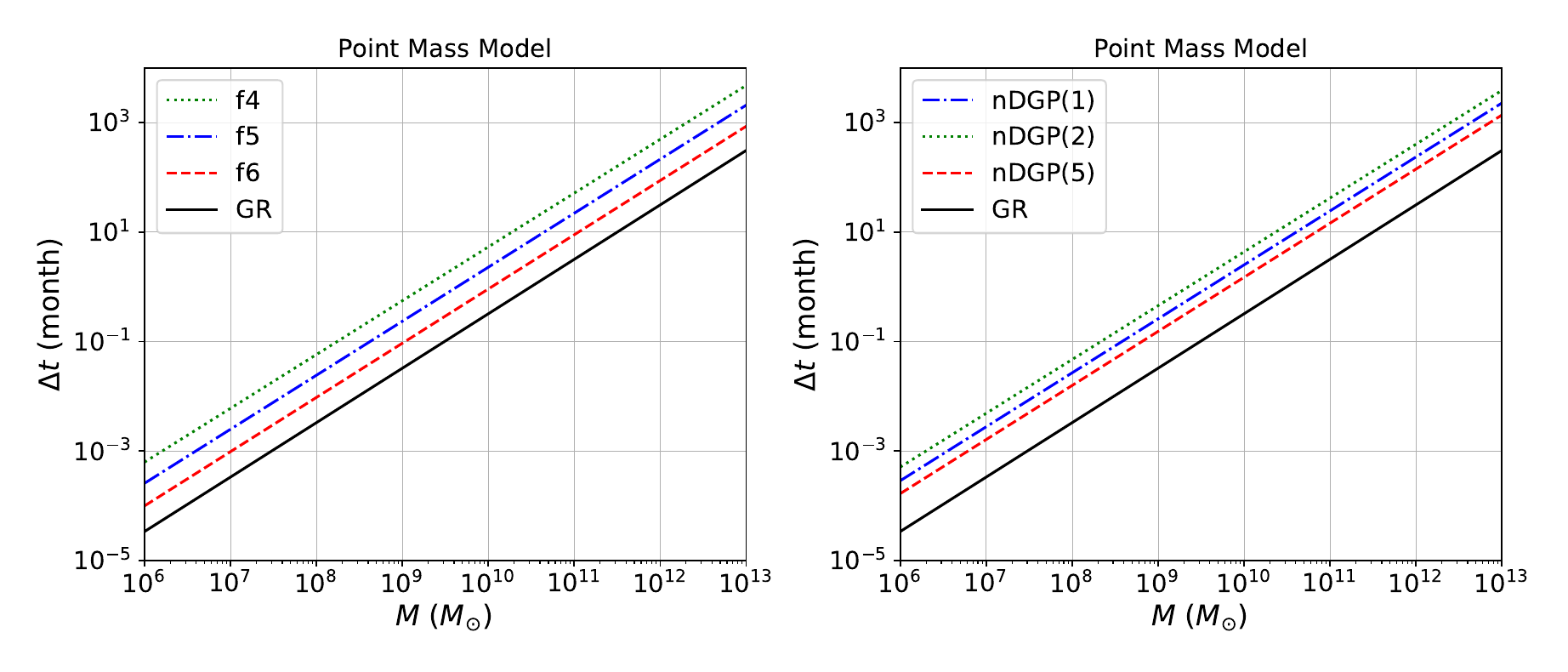}
          \includegraphics[width=0.9\linewidth]{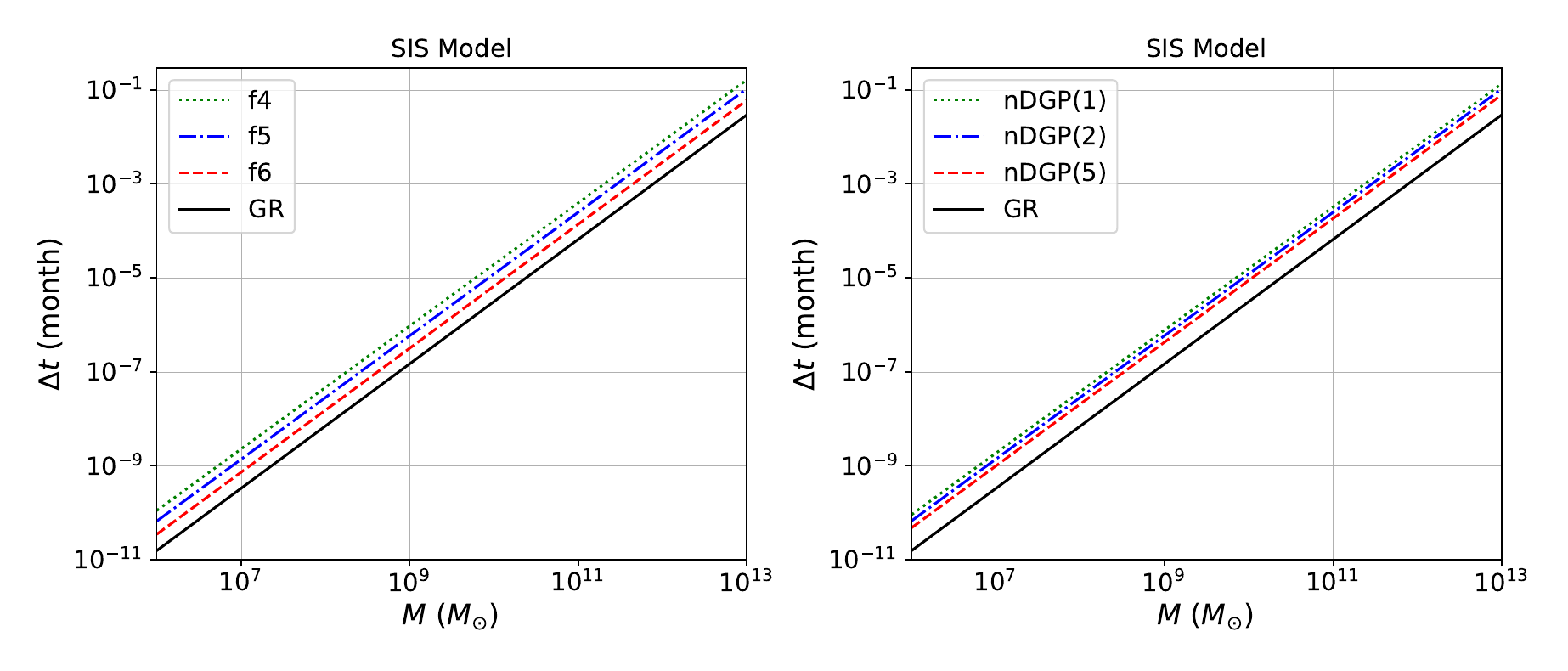}
   \caption{The lensing time delay as a function of lens mass for the Hu-Sawicki $f(R)$ (left) and nDGP (right) models, compared to findings in the standard GR framework. The analysis includes both point mass (top) and SIS (bottom) lens models.}
   \label{Fig4}
\end{figure*}

In Fig.\,\ref{Fig3}, we have depicted the lensing optical depth as a function of redshift, comparing the predictions of the Hu-Sawicki $f(R)$ and nDGP models to the GR framework.

The left panel shows the lensing optical depth for the three $f(R)$ gravity models, and that obtained from GR. Here, the $f(R)$ models, i.e., f4, f5, and f6, consistently exhibit higher lensing optical depths compared to the GR case across the entire redshift range. This suggests that the $f(R)$ gravity framework predicts a higher probability of observing gravitational lensing events than the standard GR model.

The enhanced lensing optical depth in the Hu-Sawicki $f(R)$ models can be attributed to the modified gravitational force law present in these theories. In $f(R)$ gravity, the additional scalar field degree of freedom leads to an enhancement of the gravitational force on cosmological scales, which in turn increases the deflection of light by matter structures. This stronger gravitational lensing effect is more pronounced in the f4 model, which has the largest deviation from GR among the Hu-Sawicki $f(R)$ models considered.

The right panel presents the lensing optical depth for the three nDGP models along with that for GR. Similar to the Hu-Sawicki $f(R)$ models, the nDGP models, nDGP(1), nDGP(2), and nDGP(5), exhibit higher lensing optical depths compared to GR.

The elevated lensing optical depths in the nDGP models can be attributed to the scale-dependent modifications to gravity inherent in this framework. In the nDGP model, gravity behaves normally at small scales but transitions to a stronger gravitational force on larger, cosmological scales. This scale-dependent gravitational behavior leads to more efficient deflection of light by matter structures, resulting in a higher probability of observing gravitational lensing events compared to the standard GR model.

It is important to note that the differences in lensing optical depth between the modified gravity models and GR become more pronounced at higher redshifts. This is because the deviations from GR become more significant in the earlier, more nonlinear stages of structure formation, which dominate the lensing optical depth at higher redshifts.
\subsection{Time Delay}
When light from a distant source passes near a massive object, it can experience time delays due to the gravitational field of that object. This delay occurs for two main reasons: a geometrical time delay, where the light takes different amounts of time to reach the observer depending on the path length, and the Shapiro time delay effect, caused by time dilation, which increases the time it takes light to travel a given distance from the perspective of an outside observer. The combined effect results in multiple images of the source arriving at different times, known as the time delay between images.

The time delay of a lensed light ray, compared to an undeflected ray in flat spacetime, is the difference in arrival times caused by both the curved path and the reduced speed of light due to the gravitational potential. Although the contribution from the change in path length is often considered negligible, both geometrical and potential time delays are routinely accounted for in gravitational lensing theory. These effects vary depending on the alignment of the source, lens, and observer; perfect alignment leads to maximum delays, while slight misalignments result in varying time differences. We have employed both the point mass and SIS lens models in gravitational lensing to thoroughly examine these two scenarios.

As mentioned earlier, the time delay in gravitational lensing is the difference between the travel time of a curved, lensed light ray forming one of the images and the travel time of an unlensed, straight light ray coming directly from the source. This time delay has two main components. The first is the geometric delay, which occurs because lensed light rays travel along longer, curved paths compared to the straight path of an unlensed ray, causing them to take more time to reach the observer. The second is the gravitational delay, resulting from the gravitational potential of the lensing mass. As light passes through the curved spacetime created by the lens, it experiences a time dilation effect due to gravity, further increasing the travel time of the lensed rays.

Under these conditions, specific expressions can be derived for both the point mass lens and SIS lens models. In general, the time delay equation can be expressed in the following form \cite{2019RPPh...82l6901O}
\begin{equation}
    \Delta{t_{\pm}} = \Delta{t_{\rm l}}\, \Phi (\theta_{+},\theta_{-}),
\end{equation}
where $\pm$ indices indicate the time delay between two different images that are produced by the same source. 
The typical time delay for the lens, $\Delta{t_{\rm l}}$ has the form
\begin{equation}\label{deltat}
    \Delta{t_{\rm l}} = (1+z_{\rm l}) \frac{4GM(<\theta_{\rm E})}{c^3},
\end{equation}
Here, $M(<\theta_{E})$ represents the total projected mass of the lensing object enclosed within the Einstein radius, which serves as a probe of the mass distribution. This expression assumes that the redshifts of both the lens and the source are known, and is given by
\begin{equation}\label{mtheta}
    M(<\theta_{\rm E}) = D_{\rm l}^{2} \int_0^{\theta_{\rm E}} 2\pi \theta^\prime \Sigma{(\theta^\prime)} \,d\theta^\prime = \pi D_{\rm l}^{2} \theta_{\rm E}^2 \Sigma_{\rm cr},
\end{equation}
where $\Sigma_{\rm cr}$ denotes the critical surface density, given as
\begin{equation}\label{sigmacr}
    \Sigma_{\rm cr} = \frac{c^2}{4\pi G} \frac{D_{\rm s}}{D_{\rm l} D_{\rm ls}}.
\end{equation}
By using Eqs.\,(\ref{deltat}-\ref{sigmacr}), the time delay of the lens can be clearly described as follows
\begin{equation}
    \Delta t_{\rm l} = \frac{1+z_{\rm l}}{c} \frac{D_{\rm l}D_{\rm s}}{D_{\rm ls}} \theta_{\rm E}^2.
\end{equation}

Moreover, the function $\Phi (\theta_{+},\theta_{-})$ depends on the specific mass models being considered. For a point mass lens, the expression is given by \citep{2019RPPh...82l6901O}
\begin{equation}
    \Phi (\theta_{+},\theta_{-}) = \frac{2(\theta_{+}^{2} - \theta_{-}^{2})}{(\theta_{+} + \theta_{-})^2},
\end{equation}
and for the SIS model, it can be determined as
\begin{equation}
     \Phi (\theta_{+},\theta_{-}) = \frac{(\theta_{+}^{2} - \theta_{-}^{2})}{2\theta_{+}\theta_{-}} + \ln \left(\frac{\theta_{+}}{\theta_{-}}\right).
\end{equation}

The evolution of $\Delta{t_{l}}$ depends on the specific lens models used. For a point-mass lens with perfect alignment, where $\left|\theta_{+}\right| \approx \left|\theta_{-}\right|$ and $\beta = 0$, we find that $\Phi (\theta_{+},\theta_{-}) \approx 0$. Under these conditions, the multiple images formed are symmetrically arranged. However, if the alignment is imperfect, the configuration becomes asymmetric.

In Fig.,\ref{Fig4}, we have presented the lensing time delay as a function of lens mass, considering both point mass and SIS lens models for the NFW density profile. The results provide clear insights into the impact of modified gravity models, i.e., the Hu-Sawicki $f(R)$ and nDGP, on gravitational lensing time delays compared to the standard GR framework.

The upper panels depict the lensing time delay for the point mass lens model. The time delay is plotted against the lens mass, ranging from $10^{6}$ to $10^{13} M_{\odot}$. As expected, the time delay increases monotonically with increasing lens mass, as more massive lenses have stronger gravitational potentials, leading to greater deflection of light and longer travel times for the lensed images.

The upper panels show that for the point mass lens model, the $f(R)$ gravity models (f4, f5, f6) exhibit higher lensing time delays compared to the GR model across the entire range of lens masses. Specifically, the f4 model predicts the highest time delays, followed by the f5 and f6 models, all of which are higher than the GR predictions.

Similarly, in the upper right panel, the nDGP models (nDGP(1), nDGP(2), nDGP(5)) also demonstrate higher lensing time delays compared to the GR model. The nDGP(1) model has the highest time delays, followed by nDGP(2) and nDGP(5), all exceeding the GR time delay values.

The lower panels, which depict the SIS lens model, also show a consistent trend: The f4, f5, and f6 models, as well as the nDGP(1), nDGP(2), and nDGP(5) models, exhibit higher lensing time delays across the range of lens masses compared to the GR model.

The key difference between the point mass and SIS lens models is the extent of the mass distribution. The point mass lens represents a highly localized gravitational potential, leading to stronger light bending and greater path length differences for the lensed images, which results in larger time delays. In contrast, the SIS lens has a more diffuse mass distribution, resulting in a weaker gravitational field and smaller deflections of light, thereby producing lower lensing time delays compared to the point mass model.

This observation suggests that the modified gravity theories of f(R) and nDGP predict stronger gravitational lensing effects, leading to larger time delays for the lensed images compared to the standard GR framework. The enhanced time delays in the modified gravity models can be attributed to the additional degrees of freedom introduced by these theories, which can alter the gravitational potential and the resulting light deflection.

\section{Wave Optics limit}
Thus far, we have examined our gravitational lensing system within the geometric optics regime. In the thin-lens approximation, lensing is treated as occurring within a limited region in comparison to the vast cosmological distances that waves traverse. According to this approximation, the lens mass is projected onto the lens plane, where waves propagate freely outside the lens, interacting only with a two-dimensional gravitational potential at the lens plane, which causes a sharp deflection in their path. Gravitational lensing of light is typically analyzed through the geometric optics approximation, which is applicable to most observational cases \cite{1992grle.book.....S}. However, in the case of GW lensing, longer wavelengths may challenge the validity of the geometric optics approximation. When the wavelength exceeds the Schwarzschild radius of the lens mass $M_{\rm L}$, diffraction effects become significant, leading to reduced magnification.

We consider GWs propagating under the influence of the gravitational potential of the lensing object, where the background spacetime is described by the line element \cite{2003ApJ...595.1039T}
\begin{equation}\label{ds2}
      ds^2 = g_{\mu \nu} dx^{\mu} dx^{\nu} = -\left(1+2U(r)\right)dt^2 + \left(1-2U(r)\right)dr^2,
\end{equation}
where $U(r)$ represents the Newtonian gravitational potential, considering the condition $U(r) \ll 1$. This metric describes the space-time associated with a static matter distribution under the linear approximation.

Next, we assume that the metric deviates minimally from the Minkowski metric, specifically by applying a linear perturbation to it
\begin{equation}
    g_{\mu \nu} =  g_{\mu \nu}^{(0)} + h_{\mu \nu},
\end{equation}
which is appropriate for weak gravitational fields, and $h_{\mu \nu}$ shows the metric perturbation. Under the transverse traceless Lorentz gauge condition of $h_{\mu ; \nu}^{\nu}=0$ and $h_{\mu}^{\nu}=0$, one can show that
\begin{equation}
   h_{\mu \nu ; \alpha}^{;\alpha} + 2R_{\alpha \mu \beta \nu}^{(0)} h^{\alpha \beta} = 0,
\end{equation}
where the semicolon indicates the covariant derivative with respect to $g_{\mu \nu}^{(0)}$ and $R_{\alpha \mu \beta \nu}^{(0)}$ is the background Riemann tensor. We introduce a scalar field $\phi$, which we treat instead of $h_{\mu \nu}$, propagating through the curved space-time. The propagation equation of the scalar wave is
\begin{equation}\label{partialmu}
   \partial_{\mu} (\sqrt{-g}g_{\mu \nu}\partial_{\nu}\phi).
\end{equation}
Applying Eq.\,\eqref{ds2} within Eq.\,\eqref{partialmu} yields
\begin{equation}
   (\nabla^{2}+\omega^{2})\phi = 4\omega^{2}U(\rm r)\phi.
\end{equation}
In this context, we have considered a monochromatic wave characterized by an angular frequency of $\omega=2\pi f$.

An effective approach is to define the amplification factor as
\begin{equation}
   F = \frac{\phi(f)}{\phi_{0}(f)},
\end{equation}
where $\phi_{0}(f)$ represents the wave amplitude in the absence of a gravitational potential, $U(r) = 0$. According to this expression, the amplification factor reflects the ratio of the GWs affected by lensing to those unaffected by it. Using the thin-lens approximation, one can express the diffraction integral formula for the amplification factor of gravitationally lensed waves in an expanding Universe as a function of frequency
\begin{equation}\label{amp}
   F(f) = \frac{D_{\rm s}}{D_{\rm ls}D_{\rm l}} \frac{f}{i} \int_{-\infty}^{\infty} \exp[2\pi ift_{\rm d}(\textbf{x},\textbf{y})] \,d^{2}\textbf{x},
\end{equation}
where $\textbf{x}=\xi /\xi_{0}$ and $\textbf{y}=\eta D_{\rm l} /\xi_{0}D_{\rm s} $. Here, $\eta$ represents the position vector of the source in the source plane, and $\xi$ denotes the impact parameter in the lens plane. The time delay for the arrival can then be expressed as
\begin{equation}
    t_{\rm d}(\textbf{x},\textbf{y}) = \frac{D_{\rm s}}{D_{\rm ls}D_{\rm l}} \xi_{0}^2 (1+z) \left[\frac{1}{2}\left|\textbf{x}-\textbf{y}\right|^{2} - \psi(\textbf{x}) + \phi_{\rm m}(\textbf{y}) \right],
\end{equation}
where $\psi(\textbf{x})$ is the deflection potential, $\xi_{0}$ is the characteristic length scale, and $\phi(\textbf{y})$ is known as Fermat's potential. We define the two dimensional deflection potential as
\begin{equation}
   \psi(\xi) = \frac{4G}{c^2}\int \Sigma(\xi^{\prime}) \ln\left(\frac{\left|\xi-\xi^{\prime}\right|^2}{\xi_{0}}\right) \,d^{2}\xi^{\prime}.
\end{equation}

When the signal wavelength is much smaller than the Schwarzschild radius of the lens, a simplification occurs within the framework of geometric optics. In this regime, the amplification factor can be approximated by focusing solely on the light paths that either maximize or minimize the time delay encountered by the light, i.e.,
\begin{equation}
   \nabla_{\rm x}t_{\rm d}(\textbf{x}_{\rm j},\textbf{y}) = 0.
\end{equation}
This corresponds to Fermat's principle. The integral in equation \eqref{amp} is simplified to a summation over the images
\begin{equation}
   F(f) = \sum\limits_{\rm j} \left|\mu_{\rm j}\right|^{1/2} \exp\left(2\pi id t_{\rm d,\rm j} - i\pi n_{\rm j}\right),
\end{equation}
where $\mu_{j}$ represents the magnification factor of the j-th image, which quantifies the area distortion resulting from the deflection. This distortion is determined by the determinant of the Jacobian matrix of the lens mapping, i.e., when $\theta \rightarrow \beta$
\begin{equation}
  \mu = \frac{1}{{\rm det}\left|\dfrac{\partial \beta}{\partial \theta}\right|}.
\end{equation}

\begin{figure*}
   \centering
     \includegraphics[width=0.45\linewidth]{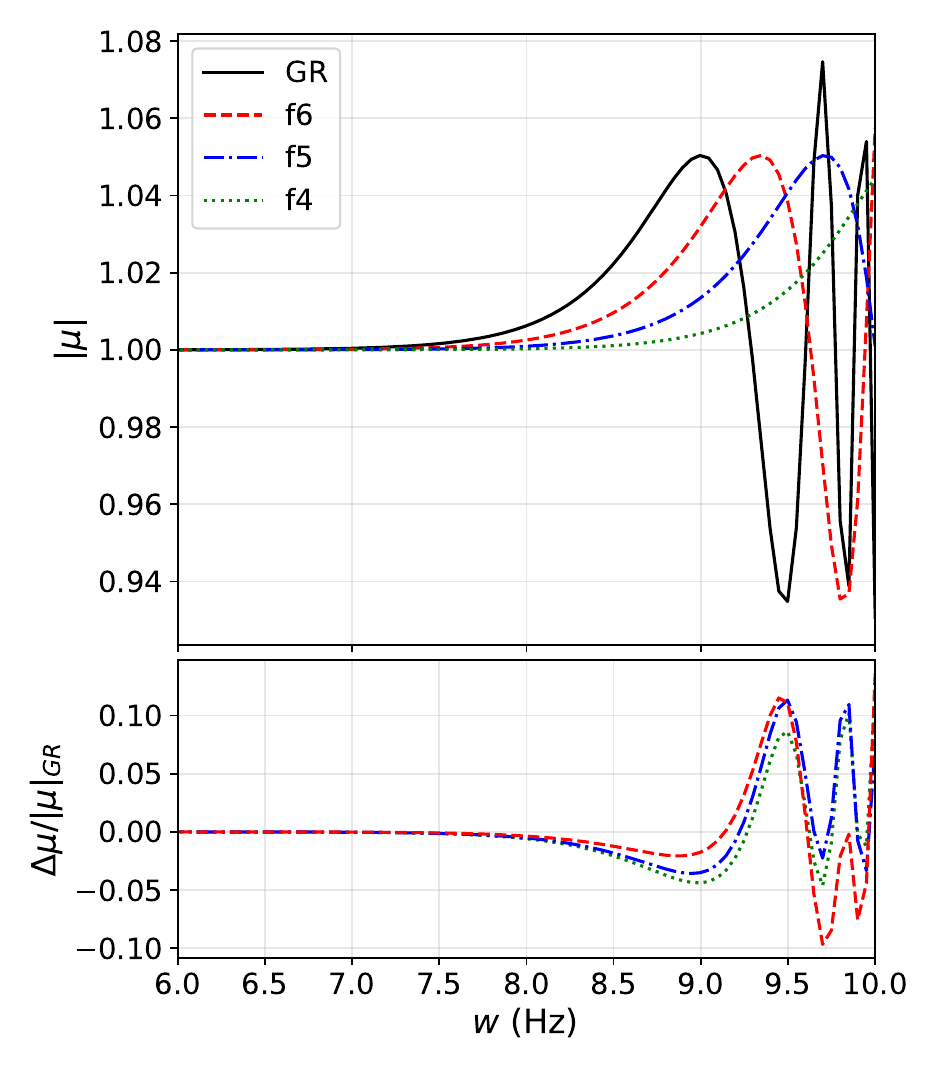}
     \includegraphics[width=0.45\linewidth]{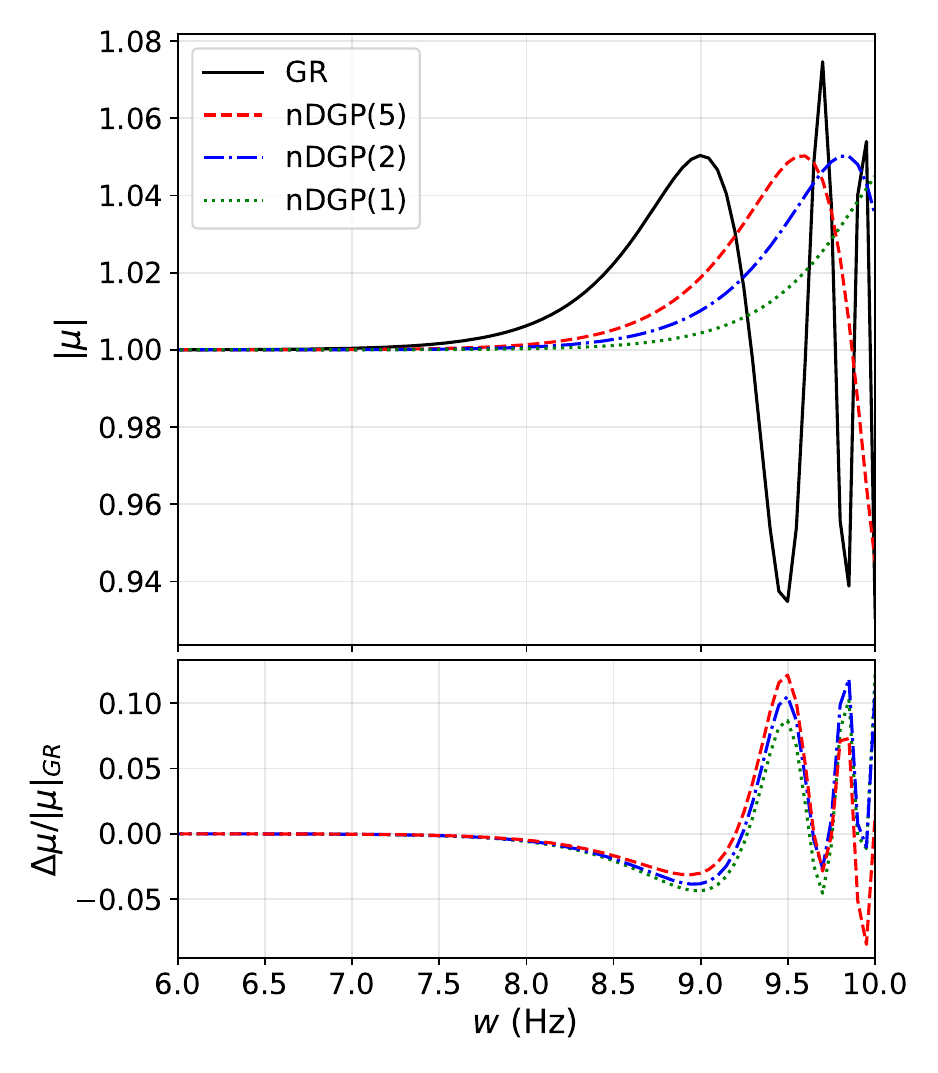}
   \caption{The magnification factors as a function of frequency for the Hu-Sawicki $f(R)$ (left) and nDGP (right) models, compared to findings in the standard GR framework. The point mass lens model and the NFW density profile are considered.}
   \label{Fig5}
\end{figure*}

\begin{figure*}
   \centering
     \includegraphics[width=0.45\linewidth]{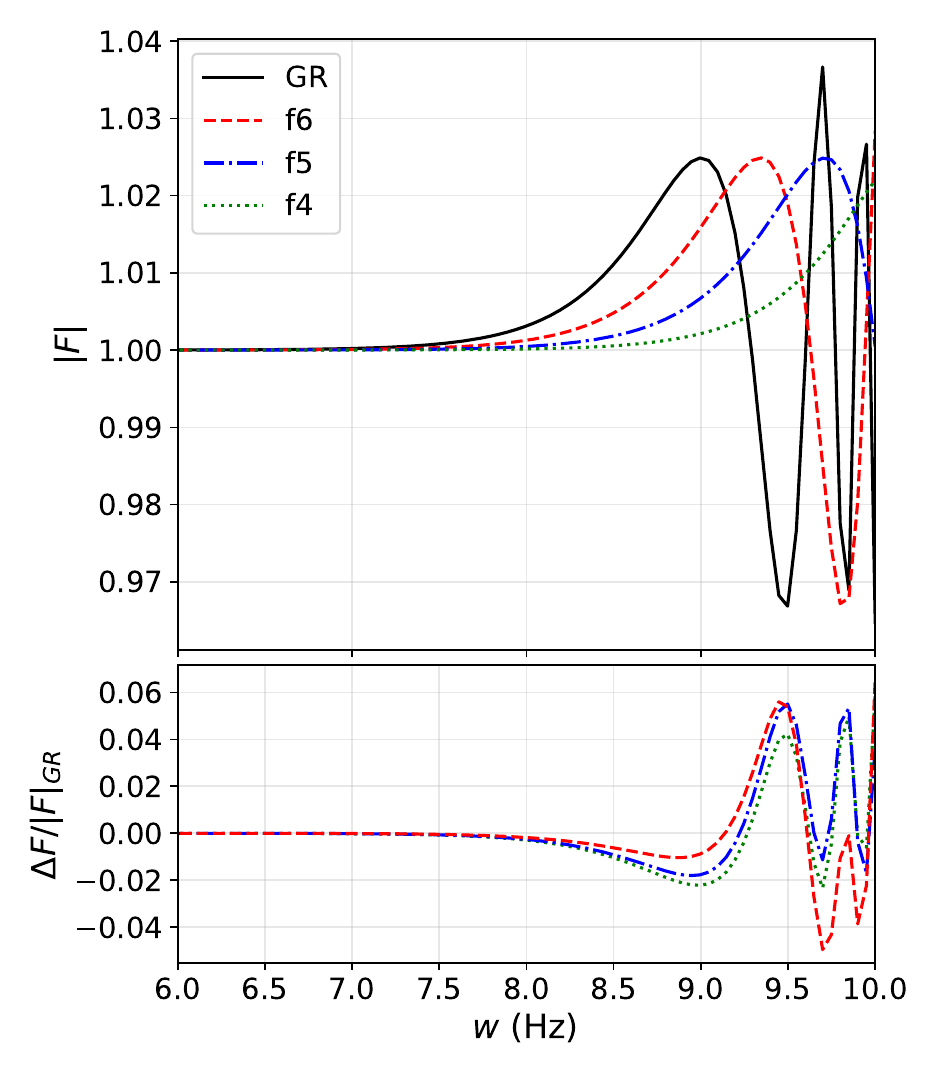}
     \includegraphics[width=0.45\linewidth]{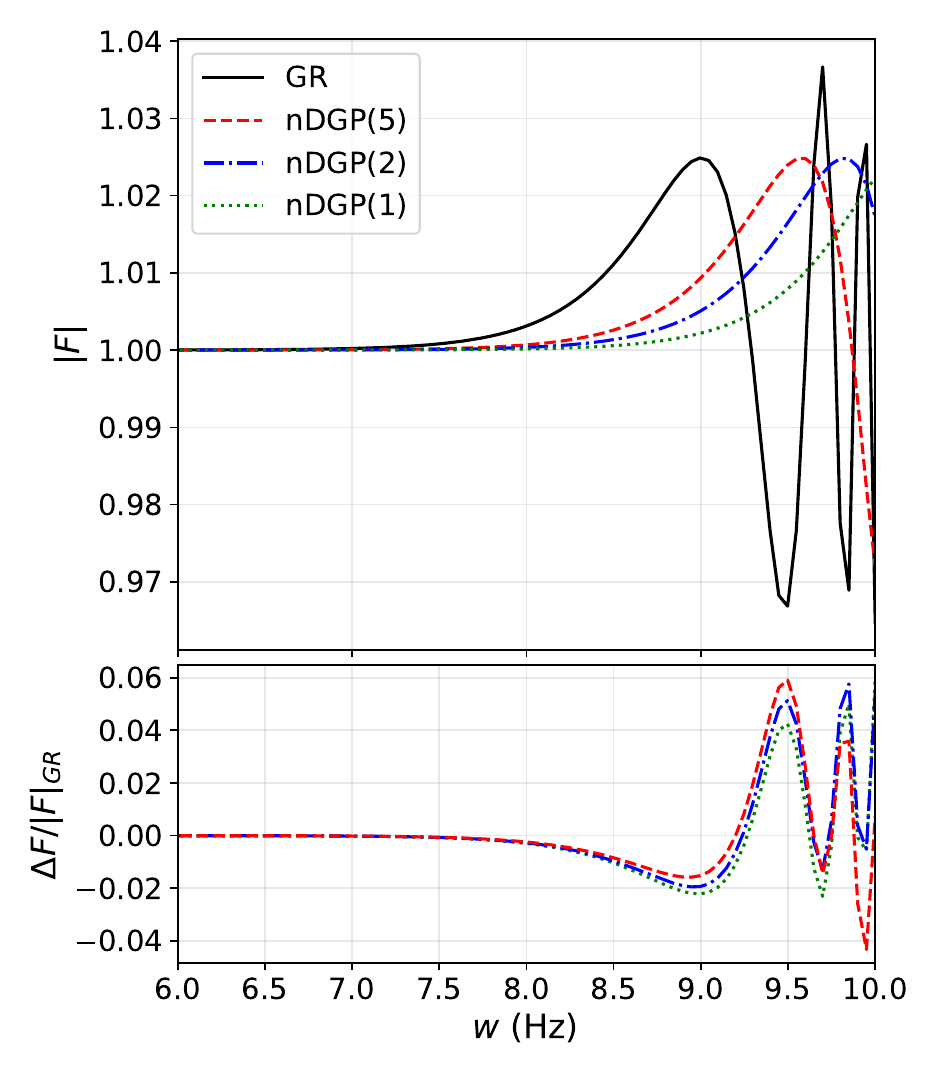}
   \caption{The amplification factors as a function of frequency for the Hu-Sawicki $f(R)$ (left) and nDGP (right) models, compared to findings in the standard GR framework. The point mass lens model and the NFW density profile are considered.}
   \label{Fig6}
\end{figure*}

After reviewing the fundamental formalism of gravitational lensing under the thin lens approximation, we now proceed to investigate this effect within various lens models outlined earlier in this paper. We begin with the case of a point mass lens. Using the Einstein angle from Eq.\,\eqref{ER1} and the non-dimensional deflection potential $\psi(x) = \ln(x)$, the amplification factor, as expressed in Eq.\,\eqref{amp}, adopts the following form:
\begin{eqnarray}\label{freq}
F(w,y) = \exp\left[\frac{i}{2}w \left(y^{2}+\log(w/2)\right)\right] \exp\left(\frac{\pi}{4}w\right) \nonumber \\ 
\quad \times \Gamma\left(1-\frac{i}{w}w\right)\,\  _{1}F_{1}\left(1-\frac{i}{w}w, 1; -\frac{i}{2}wy^2\right),
\end{eqnarray}
where $ _{1}F_{1}$ and $\Gamma$ represent the confluent hypergeometric and Gamma functions, respectively. This model introduces a dimensionless frequency $\omega$, which plays a pivotal role in characterizing wave behavior \cite{2006JCAP...01..023M}. Specifically, this parameter gauges the ratio between the Schwarzschild radius and the wavelength of the propagating wave
\begin{equation}
  w = \frac{4GM(1+z_{\rm l})}{c^3} 2\pi f \simeq 1.2\times 10^{-4} (1+z_{\rm l}) \left(\frac{M}{M_{\odot}}\right) \left(\frac{f}{1\mathrm{Hz}}\right).
\end{equation}
As $w$ approaches unity, wave effects become increasingly significant. By specifying the magnification as $\mu(w,y)=\left|F(w,y)\right|^2$, one can obtain
\begin{equation}
  \mu(w,y) = \frac{\pi w}{1-\exp(-\pi w)} \left| _{1}F_{1}\left(1-\frac{i}{2}w, 1; -\frac{i}{2}w y^2\right)\right|^2.
\end{equation}
It is important to highlight that when $y = 0$, or equivalently $\beta = 0$, the maximum magnification occurs, as this condition represents a perfect alignment between the observer, the lens, and the source. Similarly, we examine an extended source, specifically our SIS lens model. In this context, the amplification factor takes the following form:
\begin{eqnarray}
F(w,y) = \exp\left(\frac{i}{2}w y^2\right) \sum\limits_{n=0}^{\infty} \frac{\Gamma\left(1+\frac{n}{2}\right)}{n!} \hspace*{2cm}\nonumber \\
\quad \times \left[2w \exp\left(i\dfrac{3\pi}{2}\right)\right]^{n/2}\,\ _{1}F_{1}\left(1+\frac{n}{2}, 1; -\frac{i}{2}w y^2\right),
\end{eqnarray}
and the magnification is:
\begin{eqnarray}
\mu(w,y)=\bigg|\sum\limits_{n=0}^{\infty} \frac{\Gamma\left(1+\frac{n}{2}\right)}{n!}\hspace*{3.5cm}\nonumber \\
\quad \times \left[2w \exp\left(i\dfrac{3\pi}{2}\right)\right]^{n/2}\,\  _{1}F_{1}\left(1+\frac{n}{2}, 1; -\frac{i}{2}w y^2\right)\bigg|^2.
\end{eqnarray}

In Fig.\,\ref{Fig5}, we have shown the magnification factors as a function of frequency for a point mass lens considering the NFW density profile. The figure is divided into two panels, with the upper panels showing the results for the Hu-Sawicki $f(R)$ gravity models and the nDGP models compared to the standard GR framework.

In the upper left panel, the magnification factor for the $f(R)$ gravity models exhibits a wider range compared to the GR model. The f4 model, which has the largest deviation from GR among the Hu-Sawicki $f(R)$ models considered, shows the most significant difference in the magnification factor. This suggests that the modified gravitational force law present in the $f(R)$ gravity framework leads to a more pronounced deflection of light by matter structures, resulting in a broader range of magnification factors compared to the standard GR model.

The enhanced magnification in the $f(R)$ gravity models can be attributed to the additional scalar field degree of freedom introduced by these theories. This scalar field leads to a modified gravitational potential, which can result in stronger gravitational lensing effects compared to the GR scenario. The differences in the magnification factor between the $f(R)$ models and GR become more pronounced at higher frequencies, indicating that the deviations from GR become more significant in the earlier, more nonlinear stages of structure formation.

The upper right panel shows a similar trend, where the nDGP models also exhibit a wider range of magnification factors compared to the GR model. The nDGP(1) model, which has the strongest deviation from GR among the nDGP models considered, demonstrates the most significant difference in the magnification factor. The scale-dependent modifications to gravity inherent in the nDGP model are responsible for the enhanced magnification effects observed.

Interestingly, the lower panels reveal an oscillatory behavior in the deviations of the modified gravity models from the GR predictions. This oscillatory pattern suggests that the differences in the magnification factor between the modified gravity theories and GR are not monotonic but rather exhibit a more complex frequency-dependent relationship. This behavior may be related to the interplay between the modified gravitational potentials and the nature of the propagating wave, leading to interference and diffraction effects that are sensitive to the frequency of the GWs.

Fig.\,\ref{Fig6}, we have indicated the amplification factors as a function of frequency for a point mass lens considering the NFW density profile. Similar to Fig.\,\ref{Fig5}, the figure is divided into two panels, with the left panel showing the results for the Hu-Sawicki $f(R)$ gravity models (f4, f5, and f6) compared to the standard GR framework, and the right panel comparing the nDGP models (nDGP(1), nDGP(2), and nDGP(5)) to GR.

The left panel demonstrates that the amplification factors for the $f(R)$ gravity models exhibit a wider range compared to the GR model. The f4 model, which has the largest deviation from GR among the Hu-Sawicki $f(R)$ models considered, shows the most significant difference in the amplification factor. This behavior is consistent with the observations made in the left panel of Fig.\,\ref{Fig5}, where the $f(R)$ gravity models displayed a broader range of magnification factors. The enhanced amplification in the $f(R)$ gravity models can be attributed to the additional scalar field degree of freedom introduced by these theories, leading to a modified gravitational potential and stronger gravitational lensing effects.

The right panel presents a similar trend, where the nDGP models also exhibit a wider range of amplification factors compared to the GR model. The nDGP(1) model, which has the strongest deviation from GR among the nDGP models considered, demonstrates the most significant difference in the amplification factor. This is in line with the observations made in the right panel of Fig.\,\ref{Fig5}, where the nDGP models showed a broader range of magnification factors. The scale-dependent modifications to gravity inherent in the nDGP model are responsible for the enhanced amplification effects observed in this case.
\section{Conclusions}\label{sec:iv}
In this study, we have expanded on the framework of gravitational lensing by examining its behavior within modified gravity theories, specifically the Hu-Sawicki \( f(R) \) and nDGP models. Our findings have shown that these models yield distinct lensing signals compared to GR, which could significantly impact how we interpret cosmic structures and dark matter distributions. The Hu-Sawicki \( f(R) \) model, introducing a functional modification to the Ricci scalar, and the nDGP model, which involves a brane-world scenario, both produce lensing effects that may reduce the necessity of dark matter as an explanatory factor. We have found that the modified gravity models show observable differences in lensing parameters, including magnification and amplification factors in the wave optics approximation, suggesting that gravitational lensing can serve as a powerful tool for testing alternative theories of gravity and the dark sector of the Universe.

We have demonstrated that the Einstein radius varies more significantly in modified gravity models than in GR. In particular, our results indicate that the Hu-Sawicki \( f(R) \) and nDGP models yield larger angular Einstein radii, especially in high-mass halo systems, which implies stronger gravitational lensing effects in these frameworks. This enhancement can be attributed to the additional scalar field present in modified gravity models, which augments the gravitational potential at larger cosmic scales. Our findings suggest that observations of increased Einstein radii in specific lensing systems could indicate the presence of modified gravitational effects rather than just dark matter concentrations, offering a new avenue for interpreting observed cosmic structures.

In addition to Einstein radius variations, we have analyzed lensing probabilities and found that modified gravity models yield higher lensing optical depths across a range of redshifts. This increased lensing probability, especially evident in the Hu-Sawicki \( f(R) \) model, suggests that lensing events are more frequent in a modified gravity context. This trend becomes more pronounced at higher redshifts, where deviations from GR amplify. The elevated optical depth in these models aligns with the stronger gravitational potentials generated by modified gravity, implying that future high-redshift lensing observations could be an effective method for distinguishing GR from alternative gravitational models.

Furthermore, we have examined lensing time delays for both the point mass and SIS models. The modified gravity theories predict longer time delays between images compared to GR, with notable differences in high-mass lens systems. These extended time delays, resulting from enhanced gravitational potentials, could serve as an observational indicator of modified gravity effects. The SIS model results, in particular, have shown that time delays are significantly impacted by the density profile and mass distribution of the lens, suggesting that more complex gravitational fields in modified theories affect the arrival time of lensed signals more substantially than in GR.

We have also evaluated the impact of modified gravity on dark matter halo velocity distributions, showing that modified theories consistently predict broader velocity distributions. The extended range of velocities in modified gravity frameworks suggests a stronger gravitational influence on halo dynamics, which may inform our understanding of dark matter particle behavior and galaxy evolution. This insight indicates that galaxy dynamics, typically modeled under GR assumptions, might require adjustments in modified gravity contexts to accurately capture observed velocity distributions and density profiles in halos.

Additionally, we have analyzed the magnification and amplification factors under the wave optics approximation, particularly relevant for GW lensing. Our results show that both the Hu-Sawicki \( f(R) \) and nDGP models produce broader magnification and amplification factors compared to GR, with these effects varying based on frequency. In the wave optics regime, amplification factors oscillate and vary with frequency, suggesting frequency-dependent interference and diffraction effects, particularly in modified gravity models. These differences in magnification and amplification underscore the unique wave-optics-based gravitational lensing signals that modified gravity theories predict, providing another measurable avenue for testing these theories.

While our analysis has produced notable findings, future research could deepen our understanding by examining the roles of clustering, evaporation, accretion, and mergers of primordial black holes within dark matter halos. These dynamics may impact halo behavior and could strengthen gravitational lensing signals. Furthermore, incorporating additional physical elements, such as baryonic feedback mechanisms, non-thermal pressure influences, and environmental conditions, may provide a more detailed picture of dark matter halo dynamics. These enhancements have the potential to increase the precision of our analysis, reduce theoretical uncertainties, and establish a more robust framework for interpreting observations within both modified gravity and general relativity frameworks.



\bigskip
\bibliography{draft_ml}
\end{document}